# Unifying aging and frailty through complex dynamical networks

Andrew D. Rutenberg,[1] Arnold B. Mitnitski,[2] Spencer Farrell,[1] Kenneth Rockwood[2,3]

[1] Dept. of Physics and Atmospheric Science, Dalhousie University, Halifax, Nova Scotia, Canada B3H 4R2
[2] Department of Medicine, Dalhousie University, Halifax, Nova Scotia, Canada B3H 2Y9,
[3] Division of Geriatric Medicine, Dalhousie University, Halifax, Nova Scotia, Canada B3H 2E1

To explore the mechanistic relationships between aging, frailty and mortality, we developed a computational model in which possible health attributes are represented by the nodes of a complex network, with the connections showing a scale-free distribution. Each node can be either damaged (i.e. a deficit) or undamaged. Damage of connected nodes facilitates local damage and makes local recovery more difficult. Our model demonstrates the known patterns of frailty and mortality without any assumption of programmed aging. It helps us to understand how the observed maximum of the frailty index (FI) might arise. The model facilitates an initial understanding of how local damage caused by random perturbations propagates through a dynamic network of interconnected nodes. Very large model populations (here, 10 million individuals followed continuously) allow us to exploit new analytic tools, including information theory, showing, for example that highly connected nodes are more informative than less connected nodes. This model permits a better understanding of factors that influence the health trajectories of individuals.

## 1. Introduction

Aging is the cumulative effect of degradation occurring at every level of the organism. One consequence of human aging is an exponentially accelerating mortality with age, according to the Gompertz law (Kirkwood, 2015; Gavrilov and Gavrilova, 2006). This law considers age, but not health status: the potency of age as the only risk factor for mortality reflects undefined changes in health; this unmeasured heterogeneity in health (and thus in the risk of death of people of the same age) is termed "frailty" (Vaupel et al., 1979). Clinically, frailty is recognized as a multiply determined state of increased vulnerability; it also increases with age (Rockwood, 2005; Rockwood et al., 2017; Clegg et al., 2013; Xue et al., 2017). Reflecting these many determinants, a broad range of health deficits can characterize individual frailty through a frailty index (FI), which is the proportion (from 0 to 1) of possible health deficits that are present in an individual (Mitnitski et al., 2001). The FI resolves much of the otherwise unmeasured heterogeneity in health of people of the same age, and is correlated with individual mortality (Mitnitski et al., 2017; Kulminski et al., 2008; Rockwood et al., 2017; Clegg et al., 2013).

Progress in understanding frailty in humans in relation to ageing requires models. Animal models of health offer convenience, economy, and qualitatively similar behavior to human aging and mortality (Howlett, 2015). Mathematical models of aging, which have a long history (Yashin et al, 2000), can play a similar role. Computational ("in silico") models can capture individual variability of health and mortality with stochastic transitions in health states. These computational models allow us to generate large populations, examine hypotheses of cause and effect, develop new analytical tools, and explore sample size effects. Computational models of organismal ageing entail significant simplification; they are not intended to directly address particular details of individual health. Nevertheless, they can explore the mechanisms that underlie the simplicity and success of the FI (Mitnitski and Rockwood, 2015; Mitnitski et al., 2017a). How aging gives rise to frailty remains largely unspecified, requiring new approaches. Complex networks provide natural models of inter-relationships in biology, physics, and social interactions (Barabasi, 2016).

In this mini-review, we are summarizing our recent work providing a mechanistic understanding of why and how deficits accumulation summarized by the frailty index is related to aging and mortality, at the systems level of investigation, such as whole organism.

**Results and discussion**

We have used a complex network to model human aging and relate it to frailty (Figure 1) (Taneja et al., 2016; Farrell et al, 2016; Mitnitski et al, 2017b). Nodes of the network can each be either undamaged or damaged (thereby representing *deficits*); damaged nodes can also be repaired, reflecting an important source of the observed dynamics of frailty (Mitnitski et al., 2017a), also account for a possibility to recover after being damaged. Nodes correspond to generic health attributes, not explicitly identified. The connections between nodes represent significant correlative connections, which can be causal. A relatively small number of nodes (alternately, "hubs") are well connected whereas most of nodes are not, as is seen with a scale-free distribution of the number of connections (Barabasi, 2016; Taneja et al., 2016). The two most connected nodes are *mortality nodes*; a subset, being next most connected nodes which are not mortality nodes are *frailty nodes*. Frailty nodes broadly correspond to clinically or biologically significant health characteristics. Most nodes have few connections. - Nodes are damaged randomly reflecting environmental influences, intrinsic features and their interaction – such as in inflammation (Fulop et al., 2015; Jazwinski and Kim, 2017). Even so, the rate of damage of an individual node increases as more of its connected neighbors are damaged; correspondingly does the rate of recovery decrease (Taneja et al., 2016; Farrell et al, 2016; Mitnitski et al, 2017b). The overall proportion of damaged frailty nodes corresponds to the FI. The behavior of our complex network can quantitatively capture Gompertz's law, the accelerated growth of the FI with age, the broadening of the distribution of the FI with age, and its observed submaximal values (at FI<1) (Farrell et al, 2016; Mitnitski et al, 2017b).

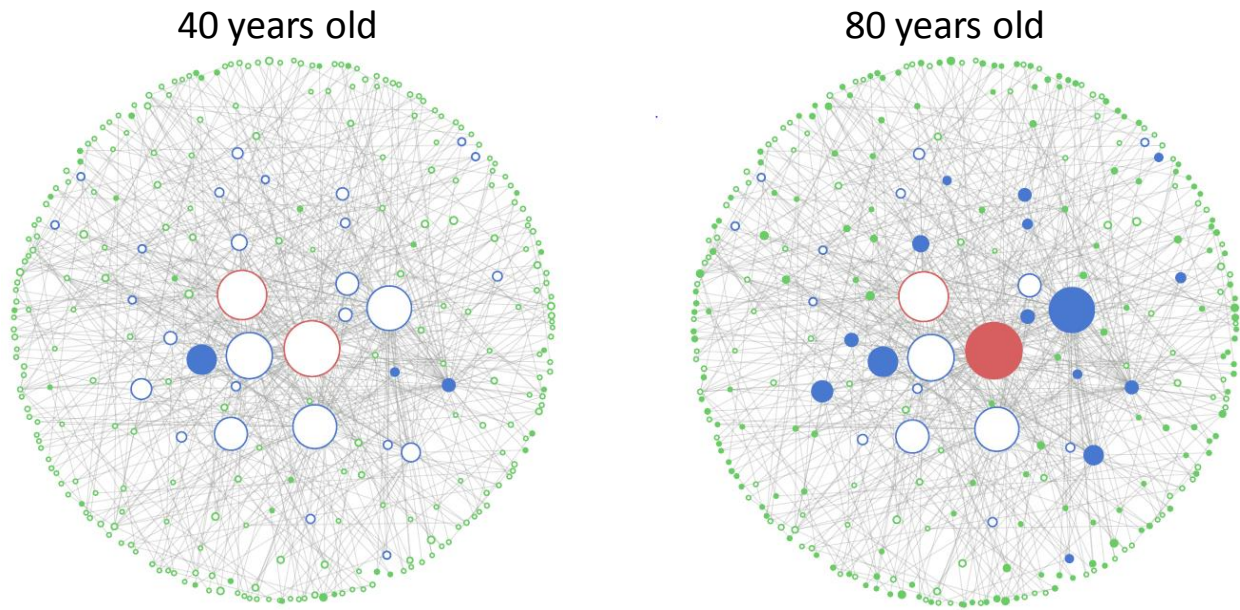

Figure 1. Connectivity networks of a model individual at age 40 (left) and then at age 80 (right). Circle size is proportional to node connectivity. Individuals die when both mortality nodes (red circles, being the two most connected nodes) are damaged. Also shown are 30 frailty nodes (blue circles), and 268 others (green circles). Damaged nodes are filled, undamaged nodes are empty. At age 40 neither mortality node is damaged, whereas 3 of 30 FI nodes are (FI=3/30=0.10) as are 34 other nodes; at age 80, one mortality node, 15 FI nodes (FI=15/30=0.50), and 173 other nodes are damaged. The individual died at age 82.

Three examples illustrate both the power and the limitations of quantitative modeling. First, a quantitative model obliges every assumption to be explicit, and this allows hypotheses of causal relationships to be explored. Even though hypotheses are difficult to falsify with a specific model, their plausibility and consistency can be validated. For example, programmed aging implies an explicit age-dependence on cellular or organismal function (Mitnitski et al., 2017b). Contrasting this is the hypothesis that aging results implicitly from the accumulation of damage (Kowald and Kirkwood, 2016). Our model supports this latter hypothesis, by showed that aging phenomenology could be recovered with no explicit age-dependent rates of damage or mortality.

Models allow us to explore quantitative hypotheses and so generate testable predictions. For our second example (Farrell et al., 2016), various observational studies have exhibited an upper frailty limit. Although many studies have a limit of about 0.7, in some it is much lower – down to 0.3 for example (Clegg et al., 2013; Drubbel et al., 2013; Harttgen et al, 2013). We were only able to recover a frailty limit below 1.0 in our model by adding a finite diagnostic sensitivity for individual deficits (Farrell et al, 2016). Since a finite sensitivity would apply to the FI in general, and not just the FI limit, we predict that studies with significantly lower limits would also have significantly lower average FI at a given age. Indeed, this is observed (Clegg et al., 2016; Drubbel et al., 2013; Harttgen et al, 2013).

In a third example of the power and limitations of modeling, consider the impact of choices about network structure. To construct our network model, we had to make assumptions about how connections were made between nodes. We assumed that as with most biological networks (Barabasi, 2016; Mitnitski et al 2017b), relatively few nodes were connected with many other nodes, whereas most nodes were only connected with a few other nodes. This echoes the intuition of "geriatric giants", that high order health impairments of, e.g., walking speed, balance, cognition, or daily function integrate information about many aspects of health; in consequence, they are highly connected. In contrast, although in biological systems no two attributes are entirely independent of each other, many physiological aspects of health are only indirectly related. We were also driven to this assumption (Figure 1), because our model did not exhibit observed aging phenomena otherwise. This implies that the network structure is important in human aging, and this is a focus of further inquiry.

We can use computational approaches to rapidly model the stochastic health trajectories and mortality of more than 10 million individuals. The model data set being clean, large, and perfectly characterized, we can explore new ways of analyzing observational data, since we can directly assess how well the analysis works. For example, information theory provides a non-parametric way of quantifying how much knowing the FI reduces our uncertainty in the mortality of an individual (Farrell et al., 2106). We can also use information measures to assess how much we learn about mortality by knowing an individual's age, or how much additional information is obtained by knowing an individual's FI given that the age is already known.

We find that larger FIs (i.e. with a larger number of actual health deficits) are most informative for younger individuals and can even exceed the information gained from knowing age alone (Farrell et al., 2016). Larger FIs indicate much earlier age-at-death than the young ages would typically indicate. On average, we also find that the information gain provided by the FI increases monotonically with the number of possible deficits included in the FI. This gives theoretical support to the recommendation to include all available health deficits in the FI. It further supports an inclusive, rather than a parsimonious, approach to evaluating the large number of potential biomarkers available through 'omics inquiries.

We can also investigate the informativeness of individual deficits. Our model demonstrates that information value of individual deficits depends on how connected they are to other nodes in the network. Deficits with more connections are more informative about mortality (Farrell et al., 2016; Mitnitski et al., 2017b). As our model network has relatively few well-connected nodes and many more less connected ones, we have a broad range of connectivities. This allows us to assess the information that individual health nodes provide about mortality, which is the focus of our current inquiries.

Age remains a convenient individual variable that provides significant information about mortality, even when the FI is known. The risk of death for older individuals is greater than for younger people with the same FI. This implies that the FI alone does not yet encapsulate the full extent of age-related damage, so that more informative FIs may be possible. Whether further improvements provide more information in addition to age remains to be seen. Since age is easy to assess, the FI complements rather than replaces age for health and mortality prediction.

Electronic health records now make possible routine FI capture in large populations using the deficits at hand (Clegg et al., 2016). Since every individual will soon have longitudinal records over their lifespan, it will be possible to include individual frailty "trajectories" into health assessment. The corresponding challenge is that opportunistic evaluation of health is likely to be biased - occurring more at times of health change or crisis than at regular intervals or annual checkups. Our computational model can precisely track when every deficit occurs for each individual, providing insights for the best use of longitudinal health data. In particular, we can quantify how sparse sampling degrades the information provided, or the effects of biased sampling that only occurs during health-changes.

Longitudinal FI analysis might be most useful when clinical intervention is being considered. Computational models allow us to characterize the effects of global or local damage to individual networks. Given that highly connected nodes are the major contributor to the risk of death, our model allows us to study how local damage to these hub nodes changes the rates of deficit accumulation and patterns of mortality. This affords investigation of how interventions to repair individual nodes might postpone damage propagation. By comparing the longitudinal behavior of the model with clinical data, our goal will be to assess the signatures of successful clinical intervention in people with complex needs.

Many important challenges remain. Our nodes are indifferent to their composition - e.g. as genetic/molecular and subcellular / cellular deficits, or damage at the level of tissue or organs, or clinically detectable deficits. It appears that these arise in that order and scale up, and hence can be detectable as biomarkers (Mitnitski et al, 2015; Lorenzi et al., 2016; Kim et al., 2017) or laboratory abnormalities (Howlett et al., 2014; Blodgett et al., 2016) or clinical deficits (Rockwood et al., 2017; Jazwinski and Kim, 2017) . Likewise, although there appear to be systematic differences in frailty by sex - at least when using self-report data, this as yet is not captured by our model (Theou et al., 2015). We have not yet found a satisfactory way to address resilience (Ukraintseva et al., 2016), although recent advances are suggestive (Gijzel et al., 2017). Nevertheless, our network approach successfully models aging phenomena.

### 3. Conclusion

Our network model of interconnected nodes reflects the interdependence of health attributes. These attributes can be summarized in the frailty index. Our recent work, reviewed here, offers solid theoretical support for how variability in deficit accumulation gives rise to variability in the risk of death for people of the same age, which is the basis of frailty in both its statistical and clinical senses. The network model shows how the local damage caused by the random perturbations propagates through the complex dynamics network of interconnected nodes. The model explains not only the known patterns of mortality (the celebrated Gompertz law) but also how health status (indicated by the frailty index) gives rise to increasing vulnerability of people when they age. Even with no age-dependent terms the model generates characteristic mortality patterns, suggesting that aging is not programmed. As with any model, there are a large number of questions still to address. With the model, however, we are now able to ask them in a way that allows explicit quantitative approaches to an area still often in thrall to semantics.

**Acknowledgement**

We thank ACENET for computational resources, along with a summer fellowship for S.F. A.D.R. thanks the Natural Sciences and Engineering Research Council of Canada (NSERC) for operating Grant No. RGPIN-2014-06245. Kenneth Rockwood receives career support from the Dalhousie Medical Research Foundation as the Kathryn Allen Weldon Professor of Alzheimer Research, and research support from the Nova Scotia Health Research Foundation and the Capital Health Research Fund and the Queen Elizabeth II Foundation Fountain Family Innovation Fund.